# Spin-the-bottle Sort and Annealing Sort: Oblivious Sorting via Round-robin Random Comparisons

Michael T. Goodrich
University of California, Irvine

**Abstract**

We study sorting algorithms based on randomized round-robin comparisons. Specifically, we study Spin-the-bottle sort, where comparisons are unrestricted, and Annealing sort, where comparisons are restricted to a distance bounded by a ***temperature*** parameter. Both algorithms are simple, randomized, data-oblivious sorting algorithms, which are useful in privacy-preserving computations, but, as we show, Annealing sort is much more efficient. We show that there is an input permutation that causes Spin-the-bottle sort to require $\Omega(n^2 \log n)$ expected time in order to succeed, and that in $O(n^2 \log n)$ time this algorithm succeeds with high probability for any input. We also show there is an implementation of Annealing sort that runs in $O(n \log n)$ time and succeeds with very high probability.

## 1 Introduction

The sorting problem is classic in computer science, with well over a fifty-year history (e.g., see [3, 20, 24, 39, 42]). In this problem, we are given an array, $A$, of $n$ elements taken from some total order and we are interested in permuting $A$ so that the elements are listed in order[1]. In this paper, we are interested in randomized sorting algorithms based on simple round-robin strategies of scanning the array $A$ while performing, for each $i = 1, 2, \ldots, n$, a compare-exchange operation between $A[i]$ and $A[s]$, where $s$ is a randomly-chosen index not equal to $i$.

In addition to its simplicity, sorting via round-robin compare-exchange operations, in this manner, is ***data-oblivious***. That is, if we view compare-exchange operations as a blackbox primitive, then the sequence of operations performed by such a randomized sorting algorithm is independent of the input permutation.

Any data-oblivious sorting algorithm can also be viewed as a ***sorting network*** [26], where the elements in the input array are provided on $n$ input wires and internal gates are compare-exchange operations. Ajtai, Komlós, and Szemerédi (AKS) [1] give a sorting network with $O(n \log n)$ compare-exchange gates, but their method is quite complicated and has a very large constant factor, even with known improvements [32, 38]. Leighton and Plaxton [27] and Goodrich [17] describe alternative randomized sorting networks that use $O(n \log n)$ compare-exchange gates and sort any given input array with very high probability. None of these previous approaches are based on simple round-robin comparison strategies, however.

Data-oblivious sorting algorithms are often motivated from their ability to be implemented in special-purpose hardware modules [24], but such algorithms also have applications in secure multi-party computation (SMC) protocols (e.g., see [4, 10, 14, 15, 28, 29]). In such protocols, two or more parties separately hold different portions of a set of data values, $\{x_1, x_2, \ldots, x_n\}$, and are interested in computing some function, $f(x_1, x_2, \ldots, x_n)$, without revealing their respective data values (e.g., see [4, 28, 40]). Thus, the design of simpler data-oblivious sorting algorithms can lead to simpler SMC protocols.

---

[1]Since we are focusing on comparison-based algorithms here, let us assume, without loss of generality, that the elements of $A$ are distinct, e.g., by a mapping $A[i] \to (A[i], i)$ and then using lexicographic ordering for comparisons.

## 1.1 Previous Related Work

In spite of their simplicity, we are not familiar with previous work on data-oblivious sorting algorithms based on round-robin random comparisons. So we review below some of the previous work on sorting that is related to the various properties that are of interest in this paper.

**Sorting via Random Comparisons.** Biedl *et al.* [5] analyze a simple algorithm, Guess-sort, which iteratively picks two elements in the input array at random and performs a compare-exchange for them, and they show that this method runs in expected time $\Theta(n^2 \log n)$. In addition, Gruber *et al.* [19] perform a more exact analysis of this algorithm, which they call Bozo-sort. Neither of these papers consider round-robin random comparisons, however.

**Quicksort.** Of course, the randomized Quicksort algorithm sorts via round-robin comparisons against a randomly-chosen element, known as a *pivot* (e.g., see [11, 18, 36]) and this leads to a sorting algorithm that runs in $O(n \log n)$ time with high probability. Even so, the set of comparisons is highly dependent on input values. Thus, randomized Quicksort is not a data-oblivious algorithm based on random round-robin compare-exchange operations.

**Shellsort.** Sorting via data-oblivious round-robin random comparisons has a similar flavor to randomized Shellsort [17], which sorts via random matchings between various subarrays of the input array. Nevertheless, there are some important differences between randomized Shellsort and sorting via round-robin random compare-exchange operations. For instance, the analysis of randomized Shellsort requires an extensive postprocessing step, which we avoid in the analysis of our randomized round-robin sorting algorithms. We also avoid the complexity of previous analyses of deterministic variants of Shellsort (e.g., see [12, 23, 33]), such as that by Pratt [34], which leads to the best known performance for deterministic Shellsort, namely, a worst-case running time of $O(n \log^2 n)$. (See also the excellent survey of Sedgewick [37].)

**Sorting via Round-robin Passes.** Sorting by deterministic round-robin passes is, of course, a classic approach, as in the well-known Bubble-sort algorithm (e.g., see [11, 18, 36]). For instance, Dobosiewicz [13] proposes sorting via various *bubble-sort passes*—doing a left-to-right sequence of compare-exchanges between elements at offset-distances apart. In addition, Incerpi and Sedgewick [21, 22] study a version of Shellsort that replaces the inner-loop with a round-robin "shaker" pass (see also [9, 41]), which is a left-to-right bubble-sort pass followed by a right-to-left bubble-sort pass. These algorithms do not ultimately lead to a time performance that is $O(n \log n)$, however.

## 1.2 Our Results

In this paper, we study two sorting algorithms based on randomized round-robin comparisons. Specifically, we study an algorithm we are calling "Spin-the-bottle sort," where comparisons in each round are arbitrary, and an algorithm we are calling "Annealing sort," where comparisons are restricted to a distance bounded by a *temperature* parameter. These algorithms are therefore similar to one another, with both being simple, data-oblivious sorting algorithms based on round-robin random compare-exchange operations.

Their respective performance is quite different, however, in that we show there is an input permutation that causes Spin-the-bottle sort to require an expected running time that is $\Omega(n^2 \log n)$ in order to succeed, and that Spin-the-bottle sort succeeds with high probability for any input permutation in $O(n^2 \log n)$ time. That is, Spin-the-bottle sort has an asymptotic expected running time that is actually *worse* than Bubble sort!

Thus, it is perhaps a bit surprising that, with just a couple of minor changes, Spin-the-bottle sort can be transformed into Annealing sort, which is much more efficient. In particular, Annealing sort is derived by applying the *simulated annealing* [25] meta-heuristic to Spin-the-bottle sort. There are, of course, multiple ways to apply this meta-heuristic, but we show there is a version of Annealing sort that runs in $O(n \log n)$ time and succeeds with very high probability[2].

---

[2] We say an algorithm succeeds **with very high probability** if success occurs with probability $1 - 1/n^\rho$, for some constant $\rho \geq 1$.

## 2  Spin-the-bottle Sort

The simplest sorting algorithm we consider in this paper is ***Spin-the-bottle*** sort[3], which is given in Figure 1.

> **while** $A$ is not sorted **do**
>    **for** $i = 1$ to $n$ **do**
>       Choose $s$ uniformly and independently at random from $\{1, 2, \ldots, i-1, i+1, \ldots, n\}$.
>       **if** ($i < s$ **and** $A[i] > A[s]$) **or** ($i > s$ **and** $A[i] < A[s]$) **then**
>          Swap $A[i]$ and $A[s]$.

Figure 1: Spin-the-bottle sort.

The test for $A$ being sorted is either done via a straightforward linear-time scan of $A$ or by a heuristic based on counting the number rounds needed until it is highly likely that $A$ is sorted. In the latter case, this leads to a data-oblivious sorting algorithm, that is, a sorting algorithm for which the sequence of comparison-exchange operations is independent of the values of the input, depending only on its size.

### 2.1  A Lower Bound on the Expected Running Time of Spin-the-bottle Sort

Our analysis of Spin-the-bottle sort is fairly straightforward and shows that this algorithm is asymptotically worse than almost all other published sorting algorithms. Nevertheless, let us go through some details of this analysis, as it provides some intuition of how improvements can be made, which in turn leads to a much more efficient algorithm, Annealing sort.

Let us begin with a lower bound on the expected running time for Spin-the-bottle sort. As was done in the analysis of Guess-Sort [5], let us consider the input array $A = (2, 1, 4, 3, \ldots, n, n-1)$, albeit now with a different argument as to why this is a difficult input instance.

This array has $N = n/2$ inversions, with each element participating in exactly one inversion. During any scan of $A$, each element that has yet to have its inversion resolved has a probability of $1/(n-1)$ of resolving its inversion. Considering the sequence of compare-exchange operations that Spin-the-bottle sort performs until $A$ is sorted, let us divide this sequence into maximal epochs of comparisons that do not resolve an inversion followed by one that does. Let $X_1, X_2, \ldots, X_N$ be a set of random variables where $X_i$ denotes the number of comparisons performed in epoch $i$, and observe that there are $N - i$ inversions remaining in $A$ after epoch $i$. Likewise, let $Y_1, Y_2, \ldots, Y_N$ be a set of random variables where $Y_i$ denotes the number of comparisons performed in epoch $i$, but only counting each comparison done such that its element, $A[i]$, has not had its inversion resolved in a previous epoch. Note that

$$X_i \geq n \left\lfloor \frac{Y_i}{n - 2(i-1)} \right\rfloor,$$

since one full round performed in epoch $i$ involves $n$ comparisons, of which $n - 2(i-1)$ are for elements that have yet to have their inversions resolved.

The running time of Spin-the-bottle sort is proportional to

$$X = \sum_{i=1}^{N} X_i.$$

---

[3]The name comes from a party game, ***Spin the bottle***, where a group of players sit in a circle and take turns, in a round-robin fashion, spinning a bottle in the middle of the circle. When it is a player's turn, he or she spins the bottle and then kisses the person of the appropriate gender nearest to where the bottle points.

Each $Y_i$ is a geometric random variable with parameter $p = 1/(n-1)$; hence, $E(Y_i) = n - 1$. Thus,

$$\begin{aligned} E(X) &= E\left(\sum_{i=1}^{N} X_i\right) \\ &\geq E\left(\sum_{i=1}^{N} n \left\lfloor \frac{Y_i}{n - 2(i-1)} \right\rfloor\right) \\ &\geq n \sum_{i=1}^{N} \left(\frac{E(Y_i)}{n - 2(i-1)} - 1\right) \\ &= n(n-1) \sum_{i=1}^{N} \frac{1}{n - 2(i-1)} - nN \\ &= n(n-1) H_{n/4}/2 - n^2/2, \end{aligned}$$

where $H_m$ denotes the $m$th Harmonic number. Thus, $E(X)$ is $\Omega(n^2 \log n)$ for this input array, giving us the following.

**Theorem 2.1:** *There is an input causing Spin-the-bottle sort to have an expected running time of $\Omega(n^2 \log n)$.*

An important lesson to take away from the proof of the above theorem is that a set of inversions between pairs of close-by elements in $A$ is sufficient to cause Spin-the-bottle sort to have a relatively large expected running time. Intuitively, the algorithm is spending a lot of time for each element $A[i]$ looking throughout the entire array for an inversion that is caused by an element right "next door" to $A[i]$. Interestingly, this same intuition applies to our upper bound for the running time of Spin-the-bottle sort.

## 2.2 An Upper Bound on the Running Time of Spin-the-bottle Sort

Let us now consider an upper bound on the running time of Spin-the-bottle sort. Our analysis is based on characterizations involving $M$, the number of inversions present in $A$ when it is given as input to the algorithm. Let $M_j$ denote the number of inversions that exist in $A$ at the beginning of round $j$ (where a round involves a complete scan of $A$), so $M_1 = M$. In addition, let $m_{i,j}$ denote the number of inversions that exist at the beginning of round $j$ and involve $A[i]$, and observe that

$$\sum_{i=1}^{n} m_{i,j} = 2M_j.$$

We divide the course of the algorithm into three phases, depending on the value of $M_j$:

- **Phase 1:** $M_j \geq 12n \log n$
- **Phase 2:** $12n \leq M_j < 12n \log n$
- **Phase 3:** $M_j < 12n$.

**Theorem 2.2:** *Given an array $A$ of $n$ elements, the three phases of Spin-the-bottle sort run in $O(n^2 \log n)$ time and sort $A$ with very high probability.*

**Proof:** See Appendix A. ∎

This, of course, is no great achievement, since there are several simple deterministic data-oblivious sorting algorithms that run in $O(n \log^2 n)$ time and even Bubble sort itself is faster than Spin-the-bottle sort, running in $O(n^2)$ time. But the above three-phase characterization nevertheless gives us some intuition that leads to a more efficient sorting algorithm, which we discuss next.

# 3 Annealing Sort

The sorting algorithm we discuss in this section is based on applying the simulated annealing [25] meta-heuristic to the sorting problem. Following an analogy from metallurgy, ***simulated annealing*** involves solving an optimization problem by a sequence of choices, such that choice $j$ is made from among some $r_j$ neighbors of a current state that are confined to be within a distance bounded from above by a parameter $T_j$ (according to an appropriate metric). Given the metallurgical analogy, the parameter $T_j$ is called the ***temperature***, which is gradually decreased during the algorithm according to an ***annealing schedule***, until it is 0, at which point the algorithm halts.

Let us apply this meta-heuristic to sorting, which is admittedly not an optimization problem, so some adaption is required. That is, let us view each round in a sorting algorithm that is similar to Spin-the-bottle sort as a step in a simulated annealing algorithm. Since each compare-exchange operation is chosen at random, let us now limit, in round $j$, the distance between candidate comparison elements to a parameter $T_j$, so as to implement the temperature metaphor, and let us also repeat the random choices for each element $r_j$ times, so as to implement a notion of neighbors of the current state under consideration. The sequence of $T_j$ and $r_j$ values defines the annealing schedule for our Annealing sort.

Formally, let us assume we are given an annealing schedule defined by the following:

- A ***temperature sequence***, $\mathcal{T} = (T_1, T_2, \ldots, T_t)$, where $T_i \geq T_{i+1}$, for $i = 1, \ldots, t-1$, and $T_t = 0$.
- A ***repetition sequence***, $\mathcal{R} = (r_1, r_2, \ldots, r_t)$, for $i = 1, \ldots, t$.

Given these two sequences, Annealing sort is as given in Figure 2.

```
for j = 1 to t do
    for i = 1 to n − 1 do
        for k = 1 to r_j do
            Let s be a random integer in the range [i + 1, min{n, i + T_j}].
            if A[i] > A[s] then
                Swap A[i] and A[s]
    for i = n downto 2 do
        for k = 1 to r_j do
            Let s be a random integer in the range [max{1, i − T_j}, i − 1].
            if A[s] > A[i] then
                Swap A[i] and A[s]
```

Figure 2: Annealing sort. It takes as input an array, $A$, of $n$ elements and an annealing schedule defined by sequences, $\mathcal{T} = (T_1, T_2, \ldots, T_t)$ and $\mathcal{R} = (r_1, r_2, \ldots, r_t)$. Note that if the compare-exchange operations are performed as a blackbox, then the algorithm is data-oblivious.

The running time of Annealing sort is $O(n \sum_{j=1}^{t} r_i)$ and its effectiveness depends on the annealing schedule, defined by $\mathcal{T} = (T_1, T_2, \ldots, T_t)$ and $\mathcal{R} = (r_1, r_2, \ldots, r_t)$. Fortunately, there is a three-phase annealing schedule that causes Annealing sort to run in $O(n \log n)$ time and succeed with very high probability:

- ***Phase 1.*** For this phase, let $\mathcal{T}_1 = (2n, 2n, n, n, n/2, n/2, n/4, n/4 \ldots, q \log^6 n, q \log^6 n)$ be the temperature sequence and let $\mathcal{R}_1 = (c, c, \ldots, c)$ be an equal-length repetition sequence (of all $c$'s), where $q \geq 1$ and $c > 1$ are constants.
- ***Phase 2.*** For this phase, let $\mathcal{T}_2 = (q \log^6 n, (q/2) \log^6 n, (q/4) \log^6 n, \ldots, g \log n)$ be the temperature sequence and let $\mathcal{R}_2 = (r, r, \ldots, r)$ be an equal-length repetition sequence, where $q$ is the constant from Phase 1, $g \geq 1$ is a constant determined in the analysis, and $r$ is $\Theta(\log n / \log \log n)$.
- ***Phase 3.*** For this phase, let $\mathcal{T}_3$ and $\mathcal{R}_3$ be sequences of length $g \log n$ of all 1's.

Given the annealing schedule defined by $\mathcal{T} = (\mathcal{T}_1, \mathcal{T}_2, \mathcal{T}_3, 0)$ and $\mathcal{R} = (\mathcal{R}_1, \mathcal{R}_2, \mathcal{R}_3, 0)$, note that the running time of Annealing sort is $O(n \log n)$. Let us therefore analyze its success probability.

## 3.1 Analysis of Phase 1

Our analysis for Phase 1 borrows some elements from our analysis of randomized Shellsort [17], as this algorithm has a somewhat similar structure of a schedule of random choices that gradually reduce in scope.

**The Probabilistic Zero-One Principle.** We begin our analysis with a probabilistic version of the *zero-one principle* (e.g., see Knuth [24]).

**Lemma 3.1 [6, 17, 35]:** *If a randomized data-oblivious sorting algorithm sorts any array of 0's and 1's of size $n$ with failure probability at most $\epsilon$, then it sorts any array of size $n$ with failure probability at most $\epsilon(n+1)$.*

This lemma is clearly only of effective use for randomized data-oblivious algorithms that have failure probabilities that are $O(n^{-\rho})$, for some constant $\rho > 1$, i.e., algorithms that succeed with very high probability.

**Shrinking Lemmas.** As we move up and down $A$ in a single pass, let us assume that we are considering the affect of this pass on an array $A$ of zeroes and ones, reasoning about how this pass impacts the ones "moving up" in $A$. We can prove a number of useful "shrinking" lemmas for the number of ones that remain in various regions (i.e., subarrays) of $A$ during this pass. (Symmetric lemmas hold for the 0's with respect to their downward movement in $A$.)

**Lemma 3.2 (Sliding-Window Lemma):** *Let $B$ be a subarray of $A$ of size $N$, and let $C$ be the subarray of $A$ of size $4N$ immediately after $B$. Suppose further there are $k \le 4\beta N$ ones in $B \cup C$, for $0 < \beta < 1$. Let $k_1^{(c)}$ be the number of ones in $B$ after a single up-and-down pass of Annealing sort with temperature $4N$ and repetition factor $c$. Then*

$$\Pr\left(k_1^{(c)} > \max\{2\beta^c N, \, 8e \log n\}\right) \le \min\{2^{-\beta^c N/2}, n^{-4}\}.$$

**Proof:** For a one to remain in a given location in $B$ it must be matched with a one in each of its $c$ compare-exchange operations in $B \cup C$ (and note that this is the extent of possibilities, since the temperature is $4N$). Moreover, we may pessimistically assume each such $c$-ary test will occur independently for each possible position in $B$ with probability at most $\beta^c$. Thus,

$$E(k_1^{(c)}) \le \beta^c N.$$

Since $k_1^{(c)}$ can, in this case, be viewed as the sum of $N$ independent 0-1 random variables, we can apply a Chernoff bound (e.g., see [30, 31]) to establish

$$\Pr\left(k_1^{(c)} > 2\beta^c N\right) \le 2^{-\beta^c N/2},$$

for the case when our bound on $E(k_1^{(c)})$ is greater than $4e \log n$. When this bound is less than or equal to $4e \log n$, we can use a Chernoff bound to establish

$$\Pr\left(k_1^{(c)} > 8e \log n\right) \le 2^{-2e \log n} \le n^{-4}.$$  ∎

**Lemma 3.3:** *Suppose we are given two regions, $B$ and $C$, of $A$, of size $N$ and $\alpha N$, respectively, for $0 < \alpha < 4$, that are contained inside a subarray of $A$ of size $4N$, with $B$ to the left of $C$, and let $k = k_1 + k_2$, where $k_1$ (resp., $k_2$) is the number of ones in $B$ (resp., $C$). Let $k_1^{(c)}$ be the number of ones in $B$ after a single up-and-down pass of Annealing sort with temperature $4N$ and repetition factor $c$. Then*

$$E\left(k_1^{(c)}\right) \leq k_1 \left(1 - \frac{\alpha}{4} + \frac{k_2}{4N}\right)^c.$$

**Proof:** A one may possibly remain in $B$ after a single (up) pass of Annealing sort with temperature $4N$, with respect to a single random choice, if it is matched with a one in $C$ or not matched with an element in $C$ at all. In a single random choice, with probability $1 - \alpha/4$, it is not matched with an element in $C$, and, if matched with an element in $C$, which occurs with probability $\alpha/4$, the probability that it is matched with a one is $k_2/(\alpha N)$. ∎

**Lemma 3.4 (Fractional-Depletion Lemma):** *Given two regions, $B$ and $C$, in $A$, of size $N$ and $\alpha N$, respectively, for $0 < \alpha < 4$, such that $B$ and $C$ are contained in a subarray of $A$ of size $4N$, with $B$ to the left of $C$, let $k = k_1 + k_2$, where $k_1$ and $k_2$ are the respective number of ones in $B$ and $C$, and suppose $k \leq 4\beta N$, for $0 < \beta < 1$. Let $k_1^{(c)}$ be the number of ones in $B$ after a single up-pass of Annealing sort with temperature $4N$ and repetition factor $c$. Then*

$$\Pr\left(k_1^{(c)} > \max\left\{2\left(1 - \frac{\alpha}{4} + \beta\right)^c N,\ 8e \log n\right\}\right) \leq \min\{2^{-(1-\alpha/4+\beta)^c N/2}, n^{-4}\}.$$

**Proof:** By Lemma 3.3, applied to this scenario,

$$E(k_1^{(c)}) \leq k_1 \left(1 - \frac{\alpha}{4} + \frac{4\beta N}{4N}\right)^c \leq \left(1 - \frac{\alpha}{4} + \beta\right)^c N.$$

Since $k_1^{(c)}$ can be viewed as the sum of $k_1$ independent 0-1 random variables, we can apply a standard Chernoff bound (e.g., see [30, 31]) to establish

$$\Pr\left(k_1^{(c)} > 2\left(1 - \frac{\alpha}{4} + \beta\right)^c N\right) \leq 2^{-(1-\alpha/4+\beta)^c N/2},$$

for the case when our bound on $E(k_1^{(c)})$ is greater than $4e \log n$. When this bound is less than or equal to $4e \log n$, we can use a Chernoff bound to establish

$$\Pr\left(k_1^{(c)} > 8e \log n\right) \leq 2^{-2e \log n} \leq n^{-4}.$$
∎

**Lemma 3.5 (Startup Lemma):** *Given two regions, $B$ and $C$, in $A$, of size $N$ and $\alpha N$, respectively, for $0 < \alpha < 4$, contained in a subarray of $A$ of size $4N$, with $B$ to the left of $C$, let $k = k_1 + k_2$, where $k_1$ and $k_2$ are the respective number of ones in $B$ and $C$, and suppose $k \leq 4\beta N$, for $0 < \beta < 1$. Let $k_1^{(c)}$ be the number of ones in $B$ after one up-pass of Annealing sort with temperature $4N$ and repetition factor $c$. Then, for any constant $\lambda > 0$ such that $1 - \alpha/4 + \beta - \lambda \leq 1 - \epsilon$, for some constant $0 < \epsilon < 1$, there is a constant $c > 1$ such that $k_1^{(c)} \leq \lambda N$, with very high probability, provided $N$ is $\Omega(\log n)$.*

**Proof:** By Lemma 3.3, so long as $k_1 \geq \lambda N$, then

$$\begin{aligned}
E(k_1^{(c)}) &\leq \left(1 - \frac{\alpha}{4} + \frac{4\beta N - \lambda N}{4N}\right)^c N \\
&\leq \left(1 - \frac{\alpha}{4} + \beta - \lambda\right)^c N \\
&\leq (1 - \epsilon)^c N.
\end{aligned}$$

Of course, we are done as soon as $k_1 \leq \lambda N$, and note that, for $c \geq \log_{1/(1-\epsilon)} \lambda/2$, we have $E(k_1^{(c)}) \leq \lambda N/2$. Thus, by a Chernoff bound, for such a constant $c$,

$$\Pr\left(k_1^{(c)} > \lambda N\right) = \Pr\left(k_1^{(c)} > 2\lambda N/2\right) \leq 2^{-\lambda N/4}.$$

The proof follows then, the fact that $N$ is $\Omega(\log n)$. ∎

Having proven the essential properties for the compare-exchange passes done in each round of Phase 1 of Annealing sort, let us now turn to the actual analysis of Phase 1.

**Bounding Dirtiness after each Iteration.** In the $2d$-th iteration of Phase 1, imagine that we partition the array $A$ into $2^d$ regions, $A_0, A_1, \ldots, A_{2^d-1}$, each of size $n/2^d$. Moreover, every two iterations with the same temperature splits a region from the previous iteration into two equal-sized halves. Thus, the algorithm can be visualized in terms of a complete binary tree, $\mathcal{B}$, with $n$ leaves. The root of $\mathcal{B}$ corresponds to a region consisting of the entire array $A$ and each leaf[4] of $\mathcal{B}$ corresponds to an individual cell, $a_i$, in $A$, of size 1. Each internal node $v$ of $\mathcal{B}$ at depth $d$ corresponds with a region, $A_i$, created in the $2d$-th iteration of the algorithm, and the children of $v$ are associated with the two regions that $A_i$ is split into during iteration $2(d+1)$.

The desired output, of course, is to have each leaf value, $a_i = 0$, for $i < n - k$, and $a_i = 1$, otherwise. We therefore refer to the transition from cell $n - k - 1$ to cell $n - k$ on the last level of $\mathcal{B}$ as the ***crossover*** point. We refer to any leaf-level region to the left of the crossover point as a ***low*** region and any leaf-level region to the right of the crossover point as a ***high*** region. We say that a region, $A_i$, corresponding to an internal node $v$ of $\mathcal{B}$, is a ***low*** region if all of $v$'s descendents are associated with low regions. Likewise, a region, $A_i$, corresponding to an internal node $v$ of , is a ***high*** region if all of $v$'s descendents are associated with high regions. Thus, we desire that low regions eventually consist of only zeroes and high regions eventually consist of only ones. A region that is neither high nor low is ***mixed***, since it is an ancestor of both low and high regions. Note that there are no mixed leaf-level regions, however.

Also note that, since Phase 1 is data-oblivious, the algorithm doesn't take any different behavior depending on whether is a region is high, low, or mixed. Nevertheless, given the shrinking lemmas presented above, we can reason about the actions of our algorithm on different regions in terms of any one of these pairs.

With each high (resp., low) region, $A_i$, define the ***dirtiness*** of $A_i$ to be the number of zeroes (resp., ones) that are present in $A_i$, that is, values of the wrong type for $A_i$. With each region, $A_i$, we associate a dirtiness bound, $\delta(A_i)$, which is a desired upper bound on the dirtiness of $A_i$. For each region, $A_i$, at depth $d$ in $\mathcal{B}$, let $j$ be the number of regions from $A_i$ to the crossover point or mixed region on that level. That is, if $A_i$ is next to the mixed region, then $j = 1$, and if $A_i$ is next to a region next to the mixed region, then $j = 2$, and so on. In general, if $A_i$ is a low leaf-level region, then $j = n - k - i - 1$, and if $A_i$ is a high leaf-level region, then $j = j - n + k$. We define the ***desired dirtiness bound***, $\delta(A_i)$, of $A_i$ as follows:

- If $j \geq 2$, then
$$\delta(A_i) = \frac{n}{2^{d+j+3}}.$$

- If $j = 1$, then
$$\delta(A_i) = \frac{n}{5 \cdot 2^d}.$$

- If $A_i$ is a mixed region, then
$$\delta(A_i) = |A_i|.$$

---
[4]This is a slight exaggeration, of course, since we terminate Phase 1 when regions have size $O(\log^6 n)$.

Thus, every mixed region trivially satisfies its desired dirtiness bound.

Because of our need for a high probability bound, we will guarantee that each region $A_i$ satisfies its desired dirtiness bound, w.v.h.p., only if $\delta(A_i) \geq 8e \log n$. If $\delta(A_i) < 8e \log n$, then we say $A_i$ is an *extreme* region, for, during our algorithm, this condition implies that $A_i$ is relatively far from the crossover point. We will show that the total dirtiness of all extreme regions is $O(\log^3 n)$ w.v.h.p. This motivates our termination of Phase 1 when the temperature is $O(\log^6 n)$.

**Lemma 3.6:** *Suppose $A_i$ is a low (resp., high) region and $\Delta$ is the cumulative dirtiness of all regions to the left (right) of $A_i$. Then any compare-exchange pass over $A$ can increase the dirtiness of $A_i$ by at most $\Delta$.*

**Proof:** If $A_i$ is a low (resp., high) region, then its dirtiness is measured by the number of ones (resp., zeroes) it contains. During any compare-exchange pass, ones can only move right, exchanging themselves with zeroes, and zeroes can only move left, exchanging themselves with ones. Thus, the only ones that can move into a low region are those to the left of it and the only zeroes that can move into a high region are those to the right of it. ∎

The inductive claim we show in Appendix B holds with very high probability is the following.

**Claim 3.7:** *After iteration $d$, for each region $A_i$, the dirtiness of $A_i$ is at most $\delta(A_i)$, provided $A_i$ is not extreme. The total dirtiness of all extreme regions is at most $8ed\log^2 n$.*

## 3.2 Analysis of Phase 2

Claim 3.7 is the essential condition we need to hold at the start of Phase 2. In this section, we analyze the degree to which Phase 2 increases the sortedness of the array $A$ further from this point.

At the beginning of Phase 2, the total dirtiness of all extreme regions is at most $8e\log^3 n$, and the size of each such region is $g\log^6 n$, for $g = 64e^2$. Without loss of generality, let us consider a one in an extreme low region. The probability that such a one fails to be compared with a zero to its right in a round of Phase 2 is at most $1/N^{1/2}$, provided $g$ is large enough. Thus, with $r = h \log n / \log \log n$, the probability such a one fails to be compared with a 0 after $r$ random comparisons at distance $N$ is at most

$$\begin{aligned}
\left(\frac{1}{N^{1/2}}\right)^{h \log n / \log \log n} &\leq \frac{1}{N^{(h/2) \log n / \log \log n}} \\
&\leq \frac{1}{(\log n)^{(h/2) \log n / \log \log n}} \\
&= \frac{1}{n^{h/2}},
\end{aligned}$$

since $N \geq \log n$ during Phase 2. Thus, with very high probability, there are no dirty extreme regions after one round of Phase 2.

Consider next a non-extreme low region that is not mixed. By Claim 3.7, the dirtiness of such a region, and all regions to its left, is, with very high probability, at most $7N/10$. Thus,

$$\begin{aligned}
E(k_1^{(r)}) &\leq \left(1 - \frac{3}{20}\right)^r N \\
&\leq e^{-(20/3)r} N.
\end{aligned}$$

Therefore, by a Chernoff bound, for $d$ and $n$ large enough,

$$\Pr\left(k_1^{(r)} > d \log N\right) \leq \frac{(eN)^{d \log N}}{\left(e^{(20/3)d \log n / \log \log n}\right)^{d \log N}}$$

$$\leq \frac{1}{e^{d\log n}}$$
$$\leq \frac{1}{n^d}.$$

Note that in the next round after this, such a region will become completely clean, w.v.h.p., since its dirtiness is below $1/N^{1/2}$ w.v.h.p.

In addition, by Lemma 3.5, since $N$ is $\Omega(\log n)$ throughout Phase 2, then, w.v.h.p, the dirtiness of regions separate from a mixed region is at most $N/6$. Thus, the above analysis applies to them as well, once they are separate from a mixed region.

Therefore, by the end of Phase 2, w.v.h.p., the only dirty regions are either mixed or within distance 2 of a mixed region. In other words, the total dirtiness of the array $A$ at the end of Phase 2 is $O(\log n)$.

### 3.3 Analysis of Phase 3

Each round of Phase 3 is guaranteed to decrease the dirtiness of $A$ by at least 1 so long as $A$ is not completely clean. This property is similar to the reason why Bubble sort works. Namely, using the zero-one principle, note that the leftmost one in $A$ will always move right until it encounters another one. Thus, a single up-pass in $A$ eliminates the leftmost one having a zero somewhere to its right. Likewise, a single down-pass in $A$ eliminates the rightmost zero having a one somewhere to its left. Thus, since the total dirtiness of $A$ is $O(\log n)$ w.v.h.p., Phase 3 will completely sort $A$ w.v.h.p.

Therefore, we have the following.

**Theorem 3.8:** *Given an array $A$ of $n$ elements, there is an annealing schedule that cause the three phases of Annealing sort to run in $O(n \log n)$ time and leave $A$ sorted with very high probability.*

## 4 Conclusion

We have given two related data-oblivious sorting algorithms based on iterated passes of round-robin random comparisons. The first, Spin-the-bottle sort requires an expected $\Omega(n^2 \log n)$ time to sort some inputs and in $O(n^2 \log n)$ time it will sort any given input sequence with very high probability. The second, Annealing sort, on the other hand, can be designed to run in $O(n \log n)$ time and sort with very high probability.

Some interesting open problems include the following.

- Our analysis is, in many ways, overly pessimistic, in order to show that Annealing sort succeeds with very high probability. Is there a simpler and shorter annealing sequence that causes Annealing sort to run in $O(n \log n)$ time and sort with very high probability?

- Both Spin-the-bottle sort and Annealing sort are highly sequential. Is there a simple[5] randomized sorting network with depth $O(\log n)$ and size $O(n \log n)$ that sorts any given input sequence with very high probability?

- Throughout this paper, we have assumed that compare-exchange operations always return the correct answer. But there are some scenarios when one would want to be tolerant of faulty compare-exchange operations (e.g., see [2, 8, 16]). Is there a version of Annealing sort that runs in $O(n \log n)$ time and sorts with high probability even if comparisons return a faulty answer uniformly at random with probability strictly less than $1/2$?

---

[5]Leighton and Plaxton [27] describe a randomized sorting network that sorts with very high probability, which is simpler than the AKS sorting network [1], but is still somewhat complicated. So the open problem would be to design a sorting network construction that is clearly simpler than the construction of Leighton and Plaxton.

**Acknowledgments**

This research was supported in part by the National Science Foundation under grants 0724806, 0713046, and 0847968, and by the Office of Naval Research under MURI grant N00014-08-1-1015.

# A  Proving the Correctness of Spin-the-bottle Sort

In this appendix, we prove Theorem 2.2, which states that, given an array $A$ of $n$ elements, the three phases of Spin-the-bottle sort run in $O(n^2 \log n)$ time and sort $A$ with very high probability.

The proof is based on showing that we can achieve each of the milestones marking each phase in $O(n^2 \log n)$ time or better.

**Phase 1.** Let $X_j$ be a random variable that equals the number of inversions resolved in round $j$ of Phase 1, and let $X_{i,j}$ denote an indicator random variable that is 1 iff we perform a comparison in iteration (round) $j$ of the algorithm between $A[i]$ and an element that caused an inversion with $A[i]$ at the beginning of round $j$. Thus,

$$X_j \geq \frac{\sum_{i=1}^n X_{i,j}}{2},$$

since each inversion involves two elements of $A$. Each of the $X_{i,j}$'s are independent. Furthermore,

$$E(X_{i,j}) = \frac{m_{i,j}}{n-1},$$

where $m_{i,j}$ denotes the number of inversions that exist at the beginning of round $j$ and involve $A[i]$. Therefore,

$$E(X_j) \geq (1/2) \sum_{i=1}^n \frac{m_{i,j}}{n-1} = M_j/(n-1),$$

where $M_j$ is the number of inversions in $A$ that exist at the beginning of round $j$. Thus, by a well-known Chernoff bound,

$$\begin{aligned}\Pr(X_j < M_j/2(n-1)) &\leq \left(\frac{e^{-1/2}}{(1/2)^{1/2}}\right)^{M_j/(n-1)} \\ &\leq 2^{-M_j/3(n-1)} \\ &\leq n^{-4},\end{aligned}$$

since we are in Phase 1. So we may assume with probability at least $1 - c/n^3$ that the following recurrence relation holds during Phase 1, for all $1 \leq j \leq cn$, for any constant $c \geq 1$:

$$M_{j+1} \leq M_j - \frac{M_j}{2n}.$$

Therefore, with probability at least $1 - 4/n^3$, there are at most $4n$ rounds during Phase 1 of Spin-the-bottle sort, since $M_1 = M < n^2$ and $M_j \geq 12n \log n$, for all $j$ during Phase 1. That is, with very high probability, Phase 1 runs in $O(n^2)$ time.

**Phase 2.** For this phase, let $X_j$ and $X_{i,j}$ denote random variables defined as in our analysis of Phase 1, with the index $j$ reset to 1 for Phase 2. In this case,

$$E(X_j) \geq M_j/(n-1) \geq 12.$$

Thus, by a similar Chernoff bound used for analyzing Phase 1,

$$\begin{aligned}\Pr(X_j < 6) &\leq \Pr(X_j < M_j/2(n-1)) \\ &\leq 2^{-M_j/3(n-1)} \\ &\leq 2^{-4},\end{aligned}$$

since we are in Phase 2. That is, with probability $1/16$ we resolve fewer than 6 inversions in round $j$ of Phase 2. Call round $j$ a *failure* in this case, and call it a *success* if it resolves at least 6 inversions. Let $Y_j$ be an indicator random variable that is 1 iff we resolve fewer than 6 inversions in round $j$ of Phase 2, or, if $j$ is larger than the number of rounds in Phase 2, then let $Y_j$ be an independent random variable that is 1 with probability $1/16$. Thus, the number of failure rounds in the first at most $4n \log n$ rounds of Phase 2 is at most

$$Y = \sum_{j=1}^{4n \log n} Y_j.$$

Note that $E(Y) = (1/4)n \log n$. Thus, by a standard Chernoff bound,

$$\begin{aligned} \Pr(Y > 2n \log n) &= \Pr(Y > 8(1/4)n \log n) \\ &\leq \left(\frac{e^7}{8^8}\right)^{(1/4)n \log n} \\ &\leq 2^{-2n \log n} \\ &= n^{-2n}. \end{aligned}$$

Note, in addition, that there can be, in total, at most $2n \log n$ successful rounds in Phase 2. Thus, with very high probability, there are only $O(n \log n)$ rounds in Phase 2. That is, with very high probability, Phase 2 runs in $O(n^2 \log n)$ time.

**Phase 3.** The analysis for this phase is similar to that for the coupon collector's problem (e.g., see [7]). At the start of this phase, there are fewer than $12n$ inversions that remain in $A$. Note that, for any such inversion, $\chi$, the probability that $\chi$ is resolved in a round of Phase 3 is at least[6] $1/n$. Let $Z_\chi^r$ be the event that $\chi$ is not resolved after $r$ rounds of Phase 3. Thus,

$$\Pr(Z_\chi^r) \leq \left(1 - \frac{1}{n}\right)^r \leq e^{-r/n}.$$

Let $R$ denote the number of rounds needed to resolve all the inversions in Phase 3. Then, for $c \geq 2$,

$$\begin{aligned} \Pr(R > cn \ln n) &\leq \Pr\left(\bigcup_\chi Z_\chi^{cn \log n}\right) \\ &\leq \sum_\chi \Pr\left(Z_\chi^{cn \log n}\right) \\ &\leq \frac{12}{n^{c-1}}. \end{aligned}$$

Thus, with very high probability, $R$ is $O(n \log n)$; hence, with very high probability, Phase 3 runs in $O(n^2 \log n)$ time. This completes the proof.

---

[6]In fact, the probability that $\chi$ is resolved in a round of Phase 3 is equal to $2/(n-1) - 1/(n-1)^2$, since each inversion has two chances of being resolved during a round.

# B  Proof of the Inductive Claim for Phase 1 of Annealing Sort

In this appendix, we prove Claim 3.7, which states that, after iteration $d$, for each region $A_i$, the dirtiness of $A_i$ is at most $\delta(A_i)$, provided $A_i$ is not extreme, and that the total dirtiness of all extreme regions is at most $8ed\log^2 n$. As mentioned above, this analysis for Phase 1 of Annealing sort borrows from our analysis of randomized Shellsort [17], as there is a similar structure to our inductive argument even though the fine details are quite different.

Let us begin at the first round, which we are viewing in terms of two regions, $A_1$ and $A_2$, of size $N = n/2$ each. Suppose that $k \le n-k$, where $k$ is the number of ones, so that $A_1$ is a low region and $A_2$ is either a high region (i.e., if $k = n-k$) or $A_2$ is mixed (the case when $k > n-k$ is symmetric). Let $k_1$ (resp., $k_2$) denote the number of ones in $A_1$ (resp., $A_2$), so $k = k_1 + k_2$. By the Startup Lemma (3.5), the dirtiness of $A_1$ will be at most $n/12$, with very high probability, since in this case (using the notation of that lemma and viewing $A$ as existing inside a larger array of size $2n$), $\alpha = 1$, $\beta \le 1/4$, and $\lambda = 1/6$, so $1 - \alpha/4 + \beta - \lambda \le 1 - 1/6$. Note that this satisfies the desired dirtiness of $A_1$, since $\delta(A_1) = n/10$ in this case. A similar argument applies to $A_2$ if it is a high region, and if $A_2$ is mixed, it trivially satisfies its desired dirtiness bound. Also, assuming $n$ is large enough, there are no extreme regions (if $n$ is so small that $A_1$ is extreme, we can immediately switch to Phase 2). The next round of Annealing sort (with temperature $2n$) can only improve the dirtiness in $A$. Thus, we satisfy the base case of our inductive argument—the dirtiness bounds for the two children of the root of $\mathcal{B}$ are satisfied with (very) high probability, and similar arguments prove the inductive claim for iterations 3 and 4, for $N = n/2^2$ and temperature $n$, and iterations 5 and 6 for $N = n/2^3$ and temperature $n/2$.

Let us now consider a general inductive step. Let us assume that, with very high probability, we have satisfied Claim 3.7 for the regions on level $d \ge 3$ and let us now consider the transition to level $d+1$, which occurs in iterations $2d+1$ and $2d+2$. In addition, we terminate this line of reasoning when the region size, $n/2^d$, becomes less than $64e^2 \log^6 n$.

**Extreme Regions.**  Let us begin with the bound for the dirtiness of extreme regions at depth $d+1$, considering the effect of iteration $2d+1$. Note that, by Lemma 3.6, regions that were extreme after iteration $2d$ will be split into regions in iteration $2d+1$ that contribute no new amounts of dirtiness to pre-existing extreme regions. That is, extreme regions get split into extreme regions. Thus, the new dirtiness for extreme regions can come only from regions that were not extreme on level $d$ of $\mathcal{B}$ that are now splitting into extreme regions on level $d+1$, which we call *freshly extreme* regions. Suppose, then, that $A_i$ is such a region, say, with a parent, $A_p$, which is $j$ regions from the mixed region on level $d$. Then the desired dirtiness bound of $A_i$'s parent region, $A_p$, is $\delta(A_p) = n/2^{d+j+3} \ge 8e\log n$, by Claim 3.7, since $A_p$ is not extreme. $A_p$ has (low-region) children, $A_i$ and $A_{i+1}$, that have desired dirtiness bounds of $\delta(A_i) = n/2^{d+1+2j+4}$ or $\delta(A_i) = n/2^{d+1+2j+3}$ and of $\delta(A_{i+1}) = n/2^{d+1+2j+3}$ or $\delta(A_{i+1}) = n/2^{d+1+2j+2}$, depending on whether the mixed region on level $d+1$ has an odd or even index. Moreover, $A_i$ (and possibly $A_{i+1}$) is freshly extreme, so $n/2^{d+1+2j+4} < 8e \log n$, which implies that $j > (\log n - d - \log \log n - 10)/2$. Nevertheless, note also that there are $O(\log n)$ new regions on this level that are just now becoming extreme, since $n/2^d > 64e^2 \log^6 n$ and $n/2^{d+j+3} \ge 8e\log n$ implies $j \le \log n - d$. So let us consider the two freshly extreme regions, $A_i$ and $A_{i+1}$, in turn, and how a pass of Annealing sort effects them (for after that they will collectively satisfy the extreme-region part of Claim 3.7).

- **Region $A_i$:** Consider the worst case for $\delta(A_i)$, namely, that $\delta(A_i) = n/2^{d+1+2j+4}$. Since $A_i$ is a left child of $A_p$, $A_i$ could get at most $n/2^{d+j+3} + 8ed\log^2 n$ ones from regions left of $A_i$, by Lemma 3.6. In addition, $A_i$ and $A_{i+1}$ could inherit at most $\delta(A_p) = n/2^{d+j+3}$ ones from $A_p$. Thus, if we let $N$ denote the size of $A_i$, i.e., $N = n/2^{d+1}$, then $A_i$ and $A_{i+1}$ together have at most $N/2^{j+1} + 3N^{1/2} \le N/2^j$ ones, since we stop Phase 1 when $N < 64e^2 \log^6 n$. In addition, assuming $j \ge 4$, regions $A_{i+2}$ and $A_{i+3}$ may inherit at most $n/2^{d+j+2}$ ones from their parent and region $A_{i+4}$ may inherit at most $n/2^{d+j+1}$ ones from its parent. Therefore, by the Sliding-Window Lemma (3.2), with $\beta = 5/2^{j+3} < 1/2^j$, the following condition holds with probability at least $1 - cn^{-4}$,

$$k_1^{(c)} \leq \max\{2\beta^c N, 8e \log n\},$$

where $k_1^{(c)}$ is the number of one left in $A_i$ after an up-pass of Annealing sort with temperature $4N$ and repetition factor $c$. Note that, if $k_1^{(c)} \leq 8e \log n$, then we have satisfied the desired dirtiness for $A_i$. Alternatively, so long as $c \geq 4$, and $j \geq 4$, then w.v.h.p.,

$$\begin{aligned} k_1^{(c)} &\leq 2\beta^c N \\ &\leq \frac{n}{2^{d+jc}} \\ &\leq \frac{n}{2^{d+1+2j+4}} \leq 8e \log n = \delta(A_i). \end{aligned}$$

- **Region $A_{i+1}$:** Consider the worst case for $\delta(A_{i+1})$, namely $\delta(A_{i+1}) = n/2^{d+1+2j+3}$. Since, in this case, $A_{i+1}$ is a right child of $A_p$, $A_{i+1}$ could get at most $n/2^{d+j+3} + 8ed \log^2 n$ ones from regions left of $A_{i+1}$, by Lemma 3.6, plus $A_{i+1}$ could inherit at most $\delta(A_p) = n/2^{d+j+3}$ ones from $A_p$ itself. In addition, since $j \geq 3$, $A_{i+2}$ and $A_{i+3}$ could inherit at most $n/2^{d+j+2}$ ones from their parent, and $A_{i+4}$ and $A_{i+5}$ could inherit at most $n/2^{d+j+1}$ ones from their parent. Thus, if we let $N$ denote the size of $A_{i+1}$, i.e., $N = n/2^{d+1}$, then $A_{i+1}$ through $A_{i+5}$ together have at most $N/2^{j+1} + 3N^{1/2} + N/2^{j+1} + N/2^j \leq 4N/2^j$ ones, since we stop Phase 1 when $N < 64e^2 \log^6 n$, and $j \geq 4$. By the Sliding-Window Lemma (3.2), applied with $\beta = 1/2^j$, the following condition holds with probability at least $1 - cn^{-4}$,

$$k_1^{(c)} \leq \max\{2\beta^c N, 8e \log n\},$$

where $k_1^{(c)}$ is the number of ones left in $A_{i+1}$ after a pass of Annealing sort with repetition factor $c$ and temperature $4N$. Note that, if $k_1^{(c)} \leq 8e \log n$, then we have satisfied the desired dirtiness bound for $A_{i+1}$. Alternatively, so long as $c \geq 4$, and $j \geq 4$, then w.v.h.p.,

$$\begin{aligned} k_1^{(c)} &\leq 2\beta^c N \\ &\leq \frac{n}{2^{d+jc}} \\ &\leq \frac{n}{2^{d+1+2j+4}} \leq 8e \log n = \delta(A_{i+1}). \end{aligned}$$

Therefore, if a low region $A_i$ or $A_{i+1}$ becomes freshly extreme in iteration $2d + 1$, then, w.v.h.p., its dirtiness is at most $8e \log n$. Since there are at most $\log n$ freshly extreme regions created in iteration $2d+1$, this implies that the total dirtiness of all extreme low regions in iteration $2d + 1$ is at most $8e(d+1) \log^2 n$, w.v.h.p., after the right-moving pass of Phase 1, by Claim 3.7. Likewise, by symmetry, a similar claim applies to the high regions after the left-moving pass of Phase 1. Moreover, by Lemma 3.6, these extreme regions will continue to satisfy Claim 3.7 after this.

**Non-extreme Regions not too Close to the Crossover Point.** Let us now consider non-extreme regions on level $d+1$ that are at least two regions away from the crossover point on level $d+1$. Consider, wlog, a low region, $A_p$, on level $d$, which is $j$ regions from the crossover point on level $d$, with $A_p$ having (low-region) children, $A_i$ and $A_{i+1}$, that have desired dirtiness bounds of $\delta(A_i) = n/2^{d+1+2j+4}$ or $\delta(A_i) = n/2^{d+1+2j+3}$ and of $\delta(A_{i+1}) = n/2^{d+1+2j+3}$ or $\delta(A_{i+1}) = n/2^{d+1+2j+2}$, depending on whether the mixed region on level $d + 1$ has an odd or even index. By Lemma 3.6, if we can show w.v.h.p. that the dirtiness of each such $A_i$ (resp., $A_{i+1}$) is at most $\delta(A_i)/3$ (resp., $\delta(A_{i+1})/3$), after the up-and-down pass of Phase 1, then no matter how many more ones come into $A_i$ or $A_{i+1}$ from the left during the rest of iteration $2d + 1$ (and $2d + 2$), they will satisfy their desired dirtiness bounds.

Let us consider the different region types (always taking the most difficult choice for each desired dirtiness in order to avoid additional cases):

- **Type 1:** $\delta(A_i) = n/2^{d+1+2j+4}$, with $j \geq 4$. Since $A_i$ is a left child of $A_p$, in this case, $A_i$ could get at most $n/2^{d+j+3} + 8ed\log^2 n$ ones from regions left of $A_i$, by Lemma 3.6. In addition, $A_i$ and $A_{i+1}$ could inherit at most $\delta(A_p) = n/2^{d+j+3}$ ones from $A_p$. Thus, if we let $N$ denote the size of $A_i$, i.e., $N = n/2^{d+1}$, then $A_i$ and $A_{i+1}$ together have at most $N/2^{j+1} + 3N^{1/2} \leq N/2^j$ ones, since we stop Phase 1 when $N < 64e^2 \log^6 n$. In addition, $A_{i+2}$ and $A_{i+3}$ inherit at most $n/2^{d+j+2}$ ones from their parent. Likewise, $A_{i+4}$ inherits at most $n/2^{d+j+1}$ ones from its parent. Thus, $A_i$ through $A_{i+4}$ inherit at most $N/2^j + N/2^{j+1} + N/2^j \leq N/2^{j-2}$ ones. Thus, we can apply the Sliding-Window Lemma (3.2), with $\beta = 1/2^j$, so that, the following condition holds with probability at least $1 - n^{-4}$, provided $c \geq 4$ and $j \geq 4$:

$$\begin{aligned} k_1^{(c)} &\leq 2\beta^c N \\ &\leq \frac{n}{2^{d+1+jc-1}} \\ &\leq \frac{n}{3 \cdot 2^{d+1+2j+4}} = \delta(A_i)/3, \end{aligned}$$

where $k_1^{(c)}$ is the number of ones left in $A_i$ after a pass of Annealing sort with repetition factor $c$.

- **Type 2:** $\delta(A_{i+1}) = n/2^{d+1+2j+3}$, with $j \geq 4$. Since $A_{i+1}$ is a right child of $A_p$, in this case, $A_{i+1}$ could get at most $n/2^{d+j+3} + 8ed\log^2 n$ ones from regions left of $A_{i+1}$, by Lemma 3.6, plus $A_{i+1}$ could inherit at most $\delta(A_p) = n/2^{d+j+3}$ ones from $A_p$. In addition, since $j > 2$, $A_{i+2}$ and $A_{i+3}$ could inherit at most $n/2^{d+j+2}$ ones from their parent. Thus, if we let $N$ denote the size of $A_{i+1}$, i.e., $N = n/2^{d+1}$, then $A_{i+1}$, $A_{i+2}$, and $A_{i+3}$ together have at most $N/2^j + 3N^{1/2} \leq N/2^{j-1}$ ones, since we stop Phase 1 when $N < 64e^2 \log^6 n$. In addition, $A_{i+4}$ and $A_{i+5}$ may inherit $n/2^{d+j+1}$ ones from their parent. Thus, $A_{i+1}$ through $A_{i+5}$ may receive $N/2^{j-1} + N/2^j \leq N/2^{j-2}$ ones. Therefore, with $\beta = 1/2^j$, we may apply the Sliding-Window Lemma (3.2) to show that, with probability at least $1 - n^{-4}$, for $j \geq 4$ and $c \geq 4$,

$$\begin{aligned} k_1^{(c)} &\leq 2\beta^c N \\ &\leq \frac{n}{2^{d+1+jc}} \\ &\leq \frac{n}{3 \cdot 2^{d+1+2j+3}} = \delta(A_{i+1})/3, \end{aligned}$$

where $k_1^{(c)}$ is the number of ones left in $A_{i+1}$ after a pass of Annealing sort with repetition factor $c$.

- **Type 3:** $\delta(A_i) = n/2^{d+1+2j+4}$, with $j = 3$. Since $A_i$ is a left child of $A_p$, in this case, $A_i$ could get at most $n/2^{d+j+3} + 8ed\log^2 n$ ones from regions left of $A_i$, by Lemma 3.6. In addition, $A_i$ and $A_{i+1}$ could inherit at most $\delta(A_p) = n/2^{d+j+3}$ ones from $A_p$. Thus, if we let $N$ denote the size of $A_i$, i.e., $N = n/2^{d+1}$, then $A_i$ and $A_{i+1}$ together have at most $N/2^{j+1} + 3N^{1/2} \leq N/2^j = N/2^3$ ones, since we stop Phase 1 when $N < 64e^2 \log^6 n$. In addition, $A_{i+2}$ and $A_{i+3}$ inherit at most $n/2^{d+j+2} = N/2^4$ ones from their parent. Finally, $A_{i+4}$ inherits at most $n/(5 \cdot 2^d) = 2N/5$ ones from its parent. Thus, $A_i$ through $A_{i+4}$ inherit at most $N/2^3 + N/2^4 + 2N/5 \leq 5N/2^3 = 5N/2^j$ ones. Thus, we can apply the Sliding-Window Lemma (3.2), with $\beta = 5/2^{j+2}$, so that, the following condition holds with probability at least $1 - n^{-4}$, for $c \geq 5$ and $j = 3$:

$$k_1^{(c)} \leq 2\beta^c N$$

$$\leq \frac{5^c n}{2^{d+(j+2)c}}$$
$$\leq \frac{n}{3 \cdot 2^{d+1+2j+4}} = \delta(A_i)/3,$$

where $k_1^{(c)}$ is the number of ones left in $A_i$ after a pass of Annealing sort with repetition factor $c$ and temperature $4N$.

- **Type 4:** $\delta(A_{i+1}) = n/2^{d+1+2j+3}$, with $j = 3$. Since $A_{i+1}$ is a right child of $A_p$, in this case, $A_{i+1}$ could get at most $n/2^{d+j+3} + 8ed \log^2 n$ ones from regions left of $A_{i+1}$, by Lemma 3.6, plus $A_{i+1}$ could inherit at most $\delta(A_p) = n/2^{d+j+3}$ ones from $A_p$. In addition, since $j > 2$, $A_{i+2}$ and $A_{i+3}$ could inherit at most $n/2^{d+j+2}$ ones from their parent. Thus, if we let $N$ denote the size of $A_{i+1}$, i.e., $N = n/2^{d+1}$, then $A_{i+1}$, $A_{i+2}$, and $A_{i+3}$ together have at most $N/2^j + 3N^{1/2} \leq N/2^{j-1}$ ones, since we stop Phase 1 when $N < 64e^2 \log^6 n$. In addition, $A_{i+4}$ and $A_{i+5}$ may inherit $n/(5 \cdot 2^d)$ ones from their parent. Thus, $A_{i+1}$ through $A_{i+5}$ may receive $N/2^{j-1} + 2N/5 < (2/3)N$ ones. Therefore, with $\beta < 1/6$, we may apply the Sliding-Window Lemma (3.2) to show that, with probability at least $1 - n^{-4}$, for $j = 3$ and $c \geq 6$,

$$k_1^{(c)} \leq 2\beta^c N$$
$$\leq \frac{n}{3^c 2^{d+1}}$$
$$\leq \frac{n}{3 \cdot 2^{d+1+2j+3}} = \delta(A_{i+1})/3,$$

where $k_1^{(c)}$ is the number of ones left in $A_{i+1}$ after a pass of Annealing sort with repetition factor $c$.

- **Type 5:** $\delta(A_i) = n/2^{d+1+2j+4}$, with $j = 2$. Since $A_i$ is a left child of $A_p$, in this case, $A_i$ could get at most $n/2^{d+j+3} + 8ed \log^2 n$ ones from regions left of $A_i$, by Lemma 3.6. In addition, $A_i$ and $A_{i+1}$ could inherit at most $\delta(A_p) = n/2^{d+j+3}$ ones from $A_p$. Thus, if we let $N$ denote the size of $A_i$, i.e., $N = n/2^{d+1}$, then $A_i$ and $A_{i+1}$ together have at most $N/2^{j+1} + 3N^{1/2} \leq N/2^j = N/2^2$ ones, since we stop Phase 1 when $N < 64e^2 \log^6 n$. In addition, $A_{i+2}$ and $A_{i+3}$ inherit at most $2N/5$ ones from their parent. Thus, we can apply the Fractional-Depletion Lemma (3.4), with $\alpha = 3$ and $\beta < 1/6$, so that the following condition holds with probability at least $1 - n^{-4}$, for $c \geq 9$ and $j = 2$:

$$k_1^{(c)} \leq 2\left(\frac{1}{4} + \frac{1}{6}\right)^c N$$
$$\leq \frac{n}{3 \cdot 2^{d+1+2j+4}} = \delta(A_i)/3,$$

where $k_1^{(c)}$ is the number of ones left in $A_i$ after a pass of Annealing sort with repetition factor $c$ and temperature $4N$.

- **Type 6:** $\delta(A_{i+1}) = n/2^{d+1+2j+3}$, with $j = 2$. Since $A_{i+1}$ is a right child of $A_p$, in this case, $A_{i+1}$ could get at most $n/2^{d+j+3} + 8ed \log^2 n$ ones from regions left of $A_{i+1}$, by Lemma 3.6, plus $A_{i+1}$ could inherit at most $\delta(A_p) = n/2^{d+j+3}$ ones from $A_p$. In addition, since $j = 2$, $A_{i+2}$ and $A_{i+3}$ could inherit at most $2N/5$ ones from their parent, where we let $N$ denote the size of $A_{i+1}$, i.e., $N = n/2^{d+1}$. Thus, $A_{i+1}$, $A_{i+2}$, and $A_{i+3}$ together have at most $N/2^{j+1} + 3N^{1/2} + 2N/5 \leq (2/3)N$ ones, since we stop Phase 1 when $N < 64e^2 \log^6 n$. Thus, $A_{i+1}$ through $A_{i+5}$ may receive $N/2^{j-1} + 2N/5 < (2/3)N$ ones. Therefore, with $\alpha = 3$ and $\beta < 1/6$, we may apply the Fractional-Depletion Lemma to show that, with probability at least $1 - n^{-4}$, for $c \geq 9$ and $j = 2$:

$$k_1^{(c)} \leq 2\left(\frac{1}{4} + \frac{1}{6}\right)^c N$$
$$\leq \frac{n}{3 \cdot 2^{d+1+2j+3}} = \delta(A_i)/3,$$

where $k_1^{(c)}$ is the number of ones left in $A_{i+1}$ after a pass of Annealing sort with repetition factor $c$ and temperature $4N$.

- **Type 7:** $\delta(A_i) = n/2^{d+1+2j+4}$, with $j = 1$. Since $A_i$ is a left child of $A_p$, in this case, $A_i$ could get at most $n/2^{d+j+2} + 8ed \log^2 n$ ones from regions left of $A_i$, by Lemma 3.6, plus $A_i$ and $A_{i+1}$ could inherit at most $\delta(A_p) = n/(5 \cdot 2^d)$ ones from $A_p$. Thus, if we let $N$ denote the size of $A_i$, i.e., $N = n/2^{d+1}$, then $A_i$ and $A_{i+1}$ together have at most $N/2^{j+1} + 2N/5 + 3N^{1/2} \le 7N/10$ ones, since we stop Phase 1 when $N < 64e^2 \log^6 n$. Thus, we may apply the Fractional-Depletion Lemma (3.4), with $\alpha = 1$ and $\beta = 0.175$, the following condition holds with probability at least $1 - n^{-4}$, for a suitably-chosen constant $c$, with $j = 1$,

$$\begin{aligned} k_1^{(c)} &\le 2(0.925)^c N \\ &\le \frac{n}{3 \cdot 2^{d+1+2j+4}} = \delta(A_i)/3, \end{aligned}$$

where $k_1^{(c)}$ is the number of ones left in $A_i$ after a pass of Annealing sort with repetition factor $c$.

Thus, $A_i$ and $A_{i+1}$ satisfy their respective desired dirtiness bounds w.v.h.p., provided they are at least two regions from the mixed region or crossover point.

**Regions near the Crossover Point.** Consider now regions near the crossover point. That is, each region with a parent that is mixed, bordering the crossover point, or next to a region that either contains or borders the crossover point. Let us focus specifically on the case when there is a mixed region on levels $d$ and $d+1$, as it is the most difficult of these scenarios.

So, having dealt with all the other regions, which have their desired dirtiness satisfied after a single up-and-down pass of Phase 1, with temperature $4N$, we are left with four regions near the crossover point, each of size $N = n/2^{d+1}$, which we will refer to as $A_1$, $A_2$, $A_3$, and $A_4$. One of $A_2$ or $A_3$ is mixed—without loss of generality, let us assume $A_3$ is mixed. At this point in the algorithm, we perform an other up-and-down pass with temperature $4N$. So, let us consider how this pass impacts the dirtiness of these four regions. Note that, by the results of the previous pass with temperature $4N$ (which were proved above), we have at this point pushed to these four regions all but at most $n/2^{d+7} + 8e(d+1) \log^2 n$ of the ones and all but at most $n/2^{d+6} + 8e(d+1) \log^2 n$ of the zeroes. Moreover, these bounds will continue to hold (and could even improve) as we perform the second up-and-down pass with temperature $4N$. Thus, at the beginning of this second pass, we know that the four regions hold between $2N - N/32 - 3N^{1/2}$ and $3N + N/64 + 3N^{1/2}$ zeroes and between $N - N/64 - 3N^{1/2}$ and $2N + N/32 + 3N^{1/2}$ ones, where $N = n/2^{d+1} > 64e^2 \log^6 n$. Let us therefore consider the impact of the second pass with temperature $4N$ for each of these four regions:

- $A_1$: this region is compared to $A_2$, $A_3$, and $A_4$, during the up-pass. Thus, we may apply the Fractional-Depletion Lemma (3.4) with $\alpha = 3$. Note, in addition, that, for $N$ large enough, since there are at most $2N + N/32 + 3N^{1/2} \le 2.2N$ ones in all of these four regions, we may apply the Fractional-Depletion Lemma with $\beta = 0.55$. Thus, the following condition holds with probability at least $1 - n^{-4}$, for a suitably-chosen constant $c$,

$$\begin{aligned} k_1^{(c)} &\le 2(0.8)^c N \\ &\le \frac{N}{32} = \delta(A_1), \end{aligned}$$

where $k_1^{(c)}$ is the number of ones left in $A_1$ after a pass of Annealing sort with repetition factor $c$ and temperature $4N$.

- $A_2$: each element of this region is compared to elements in $A_3$ and $A_4$ in the up-pass and $A_1$ in the down-pass. Note, however, that even if $A_1$ receives $N$ zeroes in the up-pass, there are still at most $2N + N/32 + 3N^{1/2} \leq 2.2N$ ones in $A_2 \cup A_3 \cup A_4$. Thus, even under this worst-case scenario (from $A_2$'s perspective), we may apply the Startup Lemma (3.5), with $\alpha = 2$, $\beta = 0.55$, and $\lambda = 1/6$, which implies that
$$1 - \alpha/4 + \beta - \lambda \leq 1 - 1/10,$$
i.e., we can take $\epsilon = 1/10$ and show that, there is a constant $c$ such that, w.v.h.p.,
$$k_1^{(c)} \leq \frac{N}{6} < \delta(A_2),$$
where $k_1^{(c)}$ is the number of ones left in $A_2$ after an up-pass of Annealing sort with repetition factor $c$ and temperature $4N$.

- $A_3$: by assumption, $A_3$ is mixed, so it automatically satisfies its desired dirtiness bound.
- $A_4$: this region is compared to $A_1$, $A_2$, and $A_3$, in the down-pass. Note further that, w.v.h.p., there are at most $3N + N/64 + 3N^{1/2} \leq 3.2N$ zeroes in these four regions, for large enough $N$. Thus, we may apply a symmetric version of the Startup Lemma (3.5), with $\alpha = 3$, $\beta = 0.8$, and $\lambda = 1/6$, which implies
$$1 - \alpha/4 + \beta - \lambda \leq 1 - 1/10,$$
i.e., we can take $\epsilon = 1/10$ and show that, there is a constant $c$ such that, w.v.h.p.,
$$k_1^{(c)} \leq \frac{N}{6} < \delta(A_4).$$
where $k_1^{(c)}$ is the number of ones left in $A_4$ after a down-pass of Annealing sort with repetition factor $c$ and temperature $4N$.

Thus, after the two up-and-down passes of Annealing sort with temperature $4N$, we will have satisfied Claim 3.7 w.v.h.p. In particular, we have proved that each region satisfies Claim 3.7 after iteration $2(d+1)$ of Phase 1 of Annealing sort with a failure probability of at most $O(n^{-4})$, for each region. Thus, since there are $O(n)$ such regions per iteration, this implies any iteration will fail with probability at most $O(n^{-3})$. Therefore, since there are $O(\log n)$ iterations, and we lose only an $O(n)$ factor in our failure probability when we apply the probabilistic zero-one principle (Lemma 3.1), when we complete the first phase of Annealing sort, w.v.h.p., at the beginning of Phase 2, the total dirtiness of all extreme regions is at most $8e \log^3 n$, and the size of each such region is $g \log^6 n$, for $g = 64e^2$.